\documentclass{article}[12pt]
\usepackage{graphics}
\begin{document}

\title{Extracting Information from the Gravitational Redshift of Compact Rotating Objects}
\author{Paul Nu\~nez\\
\small{Department of Physics,}\\
\small{University of Utah,}\\
\small{115 S 1400 E, Salt Lake City, UT 84112}\\
\small{and}\\
\small{Observatorio Astron\'omico,}\\
\small{Universidad Sergio Arboleda,}\\
\small{Cll. 74 No. 14-14, Bogot\'a, Colombia}\\
Marek Nowakowski\\
\small{Departamento de F\'isica,}
\small{Universidad de los Andes,}\\
\small{Cra. 1E No. 18A-10, Bogot\'a, Colombia.}\\
}

\maketitle

\begin{abstract}
When rotation is not taken into account, the measurement of the Gravitational Redshift can provide 
unique information about the compactness ($M/R$) of the star. Rotation alters the gravitational redshift
rendering thereby a unique determination of the compactness parameter impossible.
Nevertheless, it can be shown that by using some theoretical input, useful information on, say, the radii of
compact rotating objects can still be extracted. 
Moreover, by measuring the gravitational redshift one can infer the maximum angular velocity of the object.
As it is well known, the minimum observed periods of rotation are found in millisecond pulsars.
Here we show that millisecond
periods are actually a semi-theoretical limit that can be found by General Relativistic arguments
corresponding to the maximum angular velocity. We apply our method to compact objects such as pulsars,
white dwarfs and neutron stars.    
\end{abstract}

Keywords: maximal angular velocity; stars: neutron; stars: rotation\\
PACS: 04.40.Dg, 97.60.Jd
\section{Introduction}
Compact Objects such as neutron stars \cite{Sterg, Fried} have radii very close to their Schwarzschild radii and hence 
General Relativity should 
be used to describe the gravity close to their surface. Neglecting rotation, the geometry 
of space-time can be described using the well known spherically symmetric Schwarzschild geometry and
information on the ratio $M/R$ of a compact object can be obtained from the (observed) gravitational redshift 
(\cite{Pavlov}, \cite{Thermal}).
This is a useful, model-independent way, to gain insight into properties of neutron stars and white dwarfs.
Although it is a model-independent procedure, it relies on certain basic assumptions, like the absence of rotation and
a perfect spherical symmetry (i.e. zero quadrupole moment). 
To gain a deeper understanding it makes then sense to relax some of these assumptions and to
examine what information can be extracted from the gravitational redshift in the more general case. 
In the present paper, we
shall study the situation with a non-zero angular velocity $\Omega$, but assuming that the object is rigid enough
to allow the approximation of a negligible quadrupole moment. 
In particular, when rotation is taken into account, spherical symmetry is lost and off diagonal terms appear
in the metric which has the following general form:

\begin{equation} 
ds^2=g_{tt}\mathrm{d}t^2+g_{rr}\mathrm{d}r^2+g_{\theta \theta}\mathrm{d}\theta^2
+g_{\phi \phi}\mathrm{d}\phi^2+ 2g_{\phi t}\mathrm{d}\phi \mathrm{d}t.    \label{general metric}
\end{equation}

The exact form of this metric including magnetic field, quadrupole moment and even
radiation (\cite{Vadya}, \cite{Sanabria}, \cite{Pachon}) is still a subject of research. We will use a first and second order approximation
to this metric neglecting the effects of the quadrupole moment and the magnetic field.
Furthermore, it is expected that only relatively young Neutron Stars can have a differential rotation \cite{Sterg} 
(encountered also in boson stars \cite{boson}) which we therefore neglect here.  
\\

 Deriving the gravitational redshift from
(\ref{general metric}) we will show that, for a fixed mass, a unique solution for the radius 
of the compact object does not exist.
Instead, we obtain two different solutions which can differ by orders of magnitude as long as 
the angular velocity
is not too large. In this case, all theoretical models still favor the solution of the smaller 
radius and we can
select this solution without the necessity to refer to some details of a specific model. 
With increasing angular
momentum, the two radii approach each other and the above selection rule is not effective anymore.
However, in such a case we can define a narrow range of allowed radii which is still a valuable
model independent information for fast rotating objects.\\
 
As a bonus of our examination of the properties of the gravitational redshift 
$\mathcal{Z}$, we can show that there
exists an upper bound on the angular velocity depending on the redshift. One can even put an absolute upper
bound which is independent of $\mathcal{Z}$. Interestingly the bounds come out to 
be in the range of the observed
millisecond pulsars. This in turn allows us to assume that the angular velocity of millisecond pulsars
is indeed approaching its maximal value. The consequences of this assumption will be discussed in the text below.
One of them is the confirmation of the result that the light emitted in very fast millisecond pulsars stems mostly
from the equatorial region (\cite{Kuzmin}, \cite{Orthogonal}, \cite{Backer}). 
We compare our results to the so called `Mass Shedding Limit'
\cite{Latt-Prak-Masak} and to other work related to the maximum angular velocity. \\

One of the main results of the paper, discussed in section four, is
that our analytical formula on maximal angular
velocity is comparable to results obtained by using extensive numerical calculations.
In particular, we agree with the numerical findings of reference \cite{Haensel1}.
\\

The essential parameters which enter our equations are the radius $R$, the angular velocity $\Omega$ and the mass 
$M$. 
The minimum central density at which a neutron star is stable is simply the 
density at which neutrons become unstable to beta decay ($\rho_0\approx 8\times 10^6 g/cm^3$ \cite{Baym}). Using the
well known Oppenheimer-Volkov \cite{OV} equations and a plausible equation of state, one can construct a stellar 
model which provides the minimum mass of the neutron star to be around
$0.08 M\odot$ which is a bit unrealistic taking into account that neutron stars are remnants
of supernova explosions \cite{Zwicky}. A more realistic value for the minimum mass is of the order of $M_{\rm NS}^{\rm min}
\approx 1M\odot$ \cite{Lattimer}, which 
is actually closer the maximum mass of a white dwarf ($1.44 M\odot$).
The maximum mass of a neutron star can be found from causality arguments \cite{Rhoades}, by recalling that the speed of sound
in dense matter has to be less than the speed of light ($\frac{dp}{d\rho}\leq c^2$). This condition gives 
a maximum mass of $M^{\rm max}_{\rm NS} \approx 3M\odot$. The radii corresponding to the maximum and minimum masses can
be found using the Oppenheimer-Volkov equation. Assuming an equation of state for a degenerate neutron
fluid, the corresponding radii for the two extreme masses lie in the range $\sim 10\,km$ to 
$\sim 100 km$. This radius range can of course change if one uses a more sophisticated equation of state,
subject of current controversy \cite{Lattimer}, but the orders of magnitude remain the same \cite{Shapiro}. The 
canonical neutron star mass and radius are thought to be $\sim 1M\odot$ and $\sim 10\,km$. 
A useful observational quantity, which agrees in order of magnitude with theoretical predictions, 
is the mean ``measured'' mass of the neutron stars in a Gaussian ensemble \cite{Thorsett}, namely
\begin{equation} \label{gaussian}
\langle M \rangle_{\rm NS}=(1.35 \pm 0.05)M_{\odot}.
\end{equation}
This  value will be used if no other information on the masses is available.\\  

The paper is organized as follows. In the second section we will briefly discuss 
the gravitational redshift as emerging from a perturbative axial symmetric metric. In section three
we will use these results to determine the radius or the range of the radii of the compact object.
Section four is devoted to the maximal angular velocity derived within our approach. In section five we discuss
some improvements by taking into account more terms in the expansion of the metric. In section six
we apply our results to some chosen 
compact objects like a white dwarf and neutron stars. Finally, in section seven we present our conclusions.

\section{The Gravitational Redshift in a Perturbation Approach}
A far away observer can measure a pulsar's angular velocity $\Omega$, given by
\begin{equation}
\Omega=\frac{d\phi}{dt}=\frac{d\phi}{d\tau}\frac{d\tau}{dt}=\frac{u^{\phi}}{u^t}. \label{omega}
\end{equation}
Using (\ref{omega}), the four velocity of a stationary point on the surface can be written as 
\begin{equation}
u^{\mu}=\left( u^t,0,0,\Omega u^t\right)
\end{equation}
Through the normalization condition of the four velocity $(u_{\mu}u^{\mu}=-1)$, we obtain the time-like component
of the four velocity in terms of the metric (\ref{general metric}) and angular velocity $\Omega$:
\begin{equation}
u^t=\left(-g_{tt}-2g_{t\phi}\Omega-\Omega^2 g_{\phi\phi} \right)^{-1/2}.
\end{equation}
The redshift factor can be calculated simply by recalling that the energy of radial a photon ($\hbar=c=1$) 
is simply \cite{Carroll}
\begin{eqnarray}
\omega &=& u^{\mu}\frac{dx_{\mu}}{d\lambda}
       = u^t \left(g_{tt}\frac{dt}{d\lambda}+g_{t\phi}\frac{d\phi}{d\lambda}\right)+
           u^{\phi}\left(g_{\phi t}\frac{dt}{d\lambda}+ g_{\phi\phi}\frac{d\phi}{d\lambda}\right)\\
\nonumber
       &=& u^t E+ u^{\phi}L
       = u^t\left( E+\Omega L\right)
\end{eqnarray} 
with $\lambda$ an affine parameter and $E$ and $L$ conserved quantities due to the
existence of two killing vectors. 
As a consequence we can write for $\omega$
\begin{equation}
\omega=\frac{E+\Omega L}{\left(-g_{tt}-2g_{t\phi}\Omega-\Omega^2 g_{\phi\phi} \right)^{1/2}},
\end{equation}
and the energy perceived by a distant observer $\omega$ can be now expressed through 
\begin{equation}
\omega=\mathcal{Z}\omega_0,
\end{equation}
were $\omega_0$ is the energy at the surface and $\mathcal{Z}$ is the redshift factor. Explicitly, the
latter is given by \cite{footnote1}
\begin{equation}
\mathcal{Z}=\left(-g_{tt}-2g_{t\phi}\Omega-\Omega^2 g_{\phi\phi} \right)^{1/2}. \label{redshift}
\end{equation}
This redshift can actually be measured for many objects \cite{footnote2}. Indeed, at the end of the paper
we will employ the results of such observations.
The behavior of $\mathcal{Z}$
as a function of the radius $R$ for a fixed mass is displayed in figure 1a.\\

\begin{figure}[h]
\begin{center}
\scalebox{.6}{\includegraphics{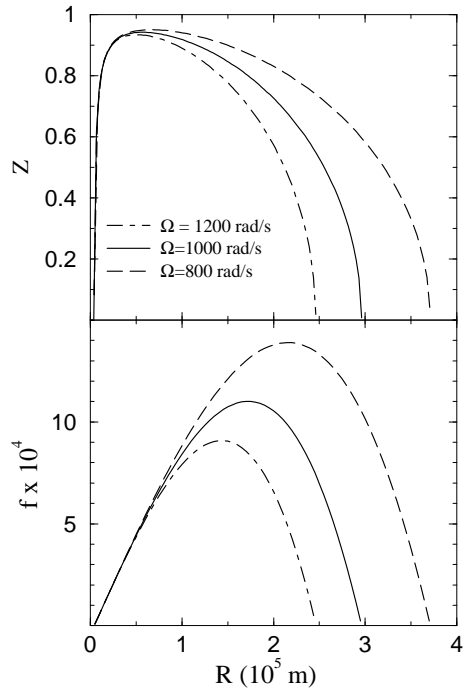}}
\caption{The top figure `a', shows the behavior of the redshift equations (\ref{redshift equation}) as
a function of the radius for different angular velocities. The bottom figure `b' shows the behavior
of the corresponding polynomial (\ref{f1}) in units of meters. In both figures $M=1.44M\odot$.  
Note how the gap between the two solutions
becomes narrow with increasing angular velocity.}

\end{center}
\end{figure}

The calculation of $\mathcal{Z}$ can be made more concrete when we consider the metric (\ref{general metric}).
To a first approximation, Zeldovich and Novikov \cite{Novikov} apply small perturbations to the 
Schwarzschild geometry. The elements of the metric  (\ref{general metric}) can be calculated 
to be \cite{Novikov}
\begin{equation}
g_{tt} = -\left( 1-\frac{2GM}{Rc^2}\right), \,\,\, 
g_{t\phi} = \frac{2GJ\sin^2{\theta}}{Rc^4}, \,\,\, 
g_{\phi\phi} = \frac{R^2\sin^2{\theta}}{c^2}. \nonumber \label{perturbative metric}
\end{equation}

Here, the condition for slow rotation is given by $J\ll MR_g c$ ($R_g$ being the Schwarzschild radius)
\cite{Novikov}.
Equation (\ref{perturbative metric}) allows us to calculate $\mathcal{Z}$ explicitly. One obtains 
\begin{equation}
\mathcal{Z}\left(M,R,\Omega,J \right)=\left( 1-\frac{2GM}{c^2 R}-\frac{4GJ\Omega \sin^2{\theta}}{c^4 R}-
\frac{R^2\Omega\sin^2{\theta}}{c^2} \right)^{1/2}, \label{J}
\end{equation}

Throughout the paper we will be using the 
Newtonian approximation for the angular momentum, ($J=\frac{2}{5}MR^2\Omega$) \cite{Mielke}. In view of the results
obtained in \cite{Shapiro2}, this is a well-based assumption violated only for extremely high angular velocities
(which are not exceeded here). 
Taking this into account, equation (\ref{J}) simplifies to
\begin{equation}
\mathcal{Z}\left(M,R,\Omega^{\prime} \right)=\left( 1-\alpha\frac{M}{R}-\beta\; M R\;\Omega^{\prime 2}-
\gamma\;R^2\Omega^{\prime 2} \right)^{1/2}, \label{redshift equation}
\end{equation}
were $\alpha$, $\beta$ and $\gamma$ are constants given by
\begin{equation}
\alpha = \frac{2G}{c^2}, \,\,\,
\beta = \frac{8G}{5 c^4}, \,\,\,
\gamma = \frac{1}{c^2} \label{constants} 
\end{equation}
and we have absorbed $\sin{\theta}$ into the angular momentum by defining
\begin{equation} \label{prime}
\Omega^{\prime} \equiv \sin{\theta} \Omega
\end{equation}
The equations (\ref{redshift}, \ref{constants}, \ref{prime}) can be now used to either solve them
for the radius by assuming a mean mass or a mass range and a measured angular velocity or, alternatively to predict
the redshift. Both ways will be used below.

\section{Determination of the Radius}
It is possible to take two different approaches when using equation (\ref{redshift equation}).
The first one is to 
demand that the term inside the parenthesis of (\ref{redshift equation}) should be greater than zero, 
such that after factoring  out $1/\sqrt{R}$ one arrives at 
\begin{equation} \label{f1}
f(0;R,M,\Omega)\equiv R-\alpha M-\beta\; M R^2\;\Omega^2-
\gamma\;R^3\Omega^2 \ge 0.
\end{equation}
The limiting values of $R$ correspond to the equal sign in the above equation. Since
this does not depend on $\mathcal{Z}$, these values have an absolute character in the sense that they give
the maximal and minimal radius for any compact object with mass $M$ and angular velocity $\Omega$ regardless 
of the value of $\mathcal{Z}$. 
Similar
reasoning applies to any other quantity derived from (\ref{f1}) (e.g. $\Omega_{\rm max}^{\prime}$ in the next section).\\
 
The behavior of the function $f(0;R,M, \Omega)$ versus $R$ is shown in figure 1b. The figure displays the global
properties of this function (which can be also inferred easily analytically), 
like the local maximum and the two zeros, one of them close to the Schwarzschild radius.\\

On the other hand, we can solve the following cubic equation for the radii
\begin{equation} \label{f2}
f(\mathcal{Z}; R, M, \Omega)=
(1-\mathcal{Z}^2)R-\alpha M-\beta\; M R^2\;\Omega^2-
\gamma\;R^3\Omega^2=0
\end{equation}
Obviously, this is is the same equation as (\ref{f1}) if we put $\mathcal{Z}$ to zero in (\ref{f2}).
Hence, we can continue  examining equation (\ref{f2}) and discuss the absolute limits by putting
$\mathcal{Z}=0$ at the end. The function $f$ with non-zero gravitational redshift has the
same global properties as (\ref{f1}).
The solutions of (\ref{f2}) can be obtained analytically by parameterizing the Cardano formulae \cite{Cardano}.
By a simple transformation one can get rid of the quadratic term in the cubic equation arriving at
$y^3 +py +q=0$. Depending on the sign of the discriminant $D=(p/3)^3 +(q/2)^2$, one can parametrize the
solution using the auxiliary variable ${\cal F} ={\rm sgn\, (q)}\sqrt{|p|/3}$. In our case $D \ge 0$ and
we parametrize the solutions through the angle $\alpha$ given by $\cos{\alpha}=q/e{\cal F}^3$
The analytical solutions are then

\begin{figure}
\begin{center}
\scalebox{.6}{\includegraphics{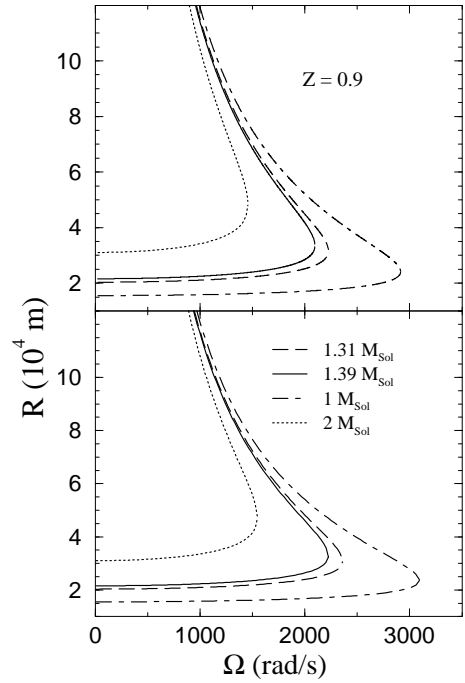}}
\caption{In these figures we show the behavior of the two solutions $r_1$  and $r_2$
  as a function of the angular velocity for different masses. The top figure
`a' corresponds to the numerical solutions using the extended metric (\ref{extended metric}), while
the bottom figure `b' corresponds to the analytical solutions  (\ref{r}, \ref{r2}) . 
Note how the solutions meet at a particular angular velocity for each mass.
}
\end{center}
\end{figure}

\begin{eqnarray}
r_1 &=& -\Delta-\mathcal{A}\cos\left\{\frac{2\pi}{3}+\frac{1}{3}\cos^{-1}{\chi}\right\} \label{r} \\ 
r_2 &=& -\Delta-\mathcal{A}\cos\left\{\frac{4\pi}{3}+\frac{1}{3}\cos^{-1}{\chi}\right\}, \label{r2}
\end{eqnarray}
were
\begin{equation}
\Delta \equiv \frac{M \beta}{3 \gamma}, \,\,\,
\mathcal{A} \equiv \frac{2}{3}\sqrt{\frac{M^2 \beta^2}{\gamma^2}+\frac{3}{\gamma \Omega^2}},
\end{equation}
and
\begin{equation}
\chi \equiv 
\frac{27\left\{  
\frac{1024}{3375}
\left(\frac{GM\Omega}{c^3}\right)^3  
+2\frac{GM\Omega}{c^3}
+\frac{8(1-\mathcal{Z}^2)}{15}\frac{GM\Omega}{c^3} 
\right\}}
{2\left\{
\frac{64}{25}   
\left(\frac{GM\Omega}{c^3} \right)^2
+3(1-\mathcal{Z}^2) 
\right\}^{3/2}}, \label{xi}
\end{equation}

In figure 2b we have plotted $r_1$ and $r_2$ versus the angular velocity $\Omega$. The upper branch
of each curve corresponds to the bigger radius meeting the lower value at some $\Omega$ (this will be discussed
in the next section in more detail). With growing $\Omega$ the difference $\Delta r$ between the two solutions becomes
smaller, however, at relative small angular velocity we can opt safely for the lower value of the
obtained radius (close to the Schwarzschild radius) as this is favored by models. Such a point of view
is not possible anymore with increasing $\Omega$.
Note also that the solutions $r_{1,2}$ do not determine a ``range'' for the radius in the strict
sense. However, since both solutions depend on the mass, a range in the mass will determine
a range for each of the two solutions. With this in mind, we can define a narrow range for the radius which will be explained 
below.\\ 

A plausible range for the mass can be given by the close
 extremes in the Gaussian distribution (\ref{gaussian}) as discussed
in the introduction. For every limiting mass we will get a curve like in figure 2, say $C_1$
for the lower mass and $C_2$ for the upper value. 
The curves with an in-between mass will fill the space between the curves $C_1$ and $C_2$.
A vertical tangent to the cusp \cite{cusp} of $C_1$ will intersect the curve $C_2$ in two points,
$r_{01}$ and $r_{02}$,  which we can take
as a definition of range of allowed radii. 
The result is a single narrow 
range which becomes smaller with increasing angular velocity and is zero at a 
 maximum $\Omega$. 
For instance, in the case of the neutron star PSR B1937+21, $r_{01}=17600\,m$ and $r_{02}=25800$ such that 
 $\Delta r=8200\, m$.
The existence of the maximum 
angular velocity corresponding to the cups of every curve allows even to sharpen this concept to 
be discussed in the next section.\\

It is of some interest to expand these solutions ($r_1$ and $r_2$)
neglecting small terms\cite{footnote3}
in $\mathcal{A}$ and $\chi$. The relevant quantities can now be approximated as
\begin{equation}
\Delta \approx \frac{8GM}{15 c^2}, \,\,\,
\mathcal{A} = \frac{2 c}{\Omega}\sqrt{\frac{1-\mathcal{Z}^2}{3}},
\end{equation}
and
\begin{equation}
\chi \approx \frac{3\sqrt{3}GM\Omega}{c^3 (1-\mathcal{Z}^2)^{3/2}}+\frac{4\sqrt{3}GM\Omega}{5c^3\sqrt{1-\mathcal{Z}^2}}
\end{equation}

When $\chi \sim 0$, one can express $r_1$ and $r_2$ as
\begin{eqnarray}
r_1 &\approx& -\Delta - \mathcal{A}\left( \frac{\sqrt{3}}{2}-\frac{\chi}{6}-\frac{\chi^2}{12\sqrt{3}}\right)\\
r_2 &\approx& -\Delta +\mathcal{A}\left(\frac{\chi}{3}+\frac{4\chi ^3}{81}\right),
\end{eqnarray}
The final results of our approximation reads
\begin{eqnarray}
r_1 &\approx& \frac{c}{\Omega}\sqrt{1-\mathcal{Z}^2}-\frac{GM}{c^2(1-\mathcal{Z}^2)}-\frac{12GM}{15c^2}
-\frac{G^2M^2 \Omega (4\mathcal{Z}^2-19)^2}{150 c^5(1-\mathcal{Z}^2)^{5/2}}\\
r_2 &\approx& \frac{2GM}{c^2(1-\mathcal{Z}^2)}+\frac{8G^3M^3\Omega^2(19-4\mathcal{Z}^2)^3}{3375 c^8(1-\mathcal{Z}^2)^4}
\end{eqnarray}

Note that the first term in $r_2$ is the Schwarzschild radius modified by a factor of
$1/(c^2(1-\mathcal{Z}^2))$. Actually, this term is the same result 
one would obtain by using a Spherically symmetric metric with no rotation.\\

It is important to remember that this approximation starts to fail when $\chi \rightarrow 1$, which occurs at 
$\mathcal{Z}\rightarrow 0$.

\section{The Limiting Angular Velocity}
To understand the origin of a maximal angular velocity it is
instructive to look at the generic behavior of the function $f$.
As already briefly mentioned, the latter will have two zeros on the positive axis and a local maximum
between them. Obviously, the case where the local maximum falls below zero is a limiting case
corresponding mathematically to $D=0$ or alternatively  to $r_1 = r_2$ (or in a yet different
method setting $\chi=1$ 
(eq. \ref{xi}) and physically corresponding to a maximally allowed angular velocity). The cusps \cite{cusp} of the curves in figures (2a and 2b) 
display this behavior.
After some algebra one obtains
\begin{eqnarray}
\Omega_{max}^{\prime}(\mathcal{Z})=\frac{c^3}{32 GM}\sqrt{\frac{5}{2}}\biggl\{
\!\!\!&-&\!\!\! 675-360(1-\mathcal{Z}^2)+16(1-\mathcal{Z}^2)^2
\nonumber \\
\!\!\!&+&\!\!\! \left(
(5+4(1-\mathcal{Z}^2))(45+4(1-\mathcal{Z}^2))^3\right)^{1/2}\biggr\}^{1/2}.
\label{maximum angular velocity}
\end{eqnarray}
As described above, we can obtain the `absolute' value of $\Omega_{max}$ by setting
$\mathcal{Z}=0$.
This is the `absolute' upper bound on the angular velocity which turns out to be
\begin{equation} 
\Omega_{max}^{\prime (1)}=\frac{5c^3}{32 G M}.\label{absolute max}
\end{equation}
A numerical value can be found by taking $M= 1.35M_{\odot}$, so that
\begin{equation} \label{Omax1}
\Omega_{max} ^{\prime}\approx 2.35 \times 10^4\, {\rm rad}/s
\end{equation}
is only one order of magnitude away from the observed millisecond pulsars. 
When $\mathcal{Z}$ is different from zero, $\Omega_{max}$ can be consistent with the largest 
angular velocities observed in millisecond 
pulsars\cite{footnote4}. The result for typical redshifts around $\mathcal{Z}\sim 1$ is
\begin{equation} \label{Omax2}
\Omega_{max}^{\prime}(\mathcal{Z})\approx \frac{c^3 }{GM}\left(\frac{2}{3}(1-\mathcal{Z})\right)^{3/2}
\end{equation}
One can interpret the equation (\ref{Omax2}) in two different, but related ways.
Both ways have to do with the evidence that milli-second pulsars are orthogonal rotators
($\theta=\pi/2$)
\cite{Kuzmin, Orthogonal, Backer}.
Since the right hand side of (\ref{Omax2}) agrees already with the angular
velocity of fast spinning objects, the emission angle $\theta$ must be close to $\pi/2$ (see equation 
(\ref{prime})). As we shall see below, this result can in turn be used to learn about the orientation of 
the magnetic axis in rapidly rotating objects, particularly neutron stars.\\

The standard model for the pulsar emission mechanism was developed independently by Pacini \cite{Pacini}
and Gold \cite{Gold} (see also 'lighthouse' model \cite{Lorimer}), 
and will now be described briefly. Since the rotation axis 
is not aligned with the magnetic axis, a changing magnetic field
will induce electric fields at the magnetic poles in for example, a neutron star. These electric fields
will eject particles which will follow helicoidal paths around the magnetic field lines. The ejected 
particles will in turn, emit a narrow cone ($\sim 10^{\circ}$ \cite{Shapiro}) of radiation parallel to the magnetic axis. 
From this argument it can be inferred that the angle between the
magnetic axis and the rotation axis is approximately the same as the emission angle $\theta$.\\

\begin{figure}[h]
\begin{center}
\scalebox{.6}{
\includegraphics{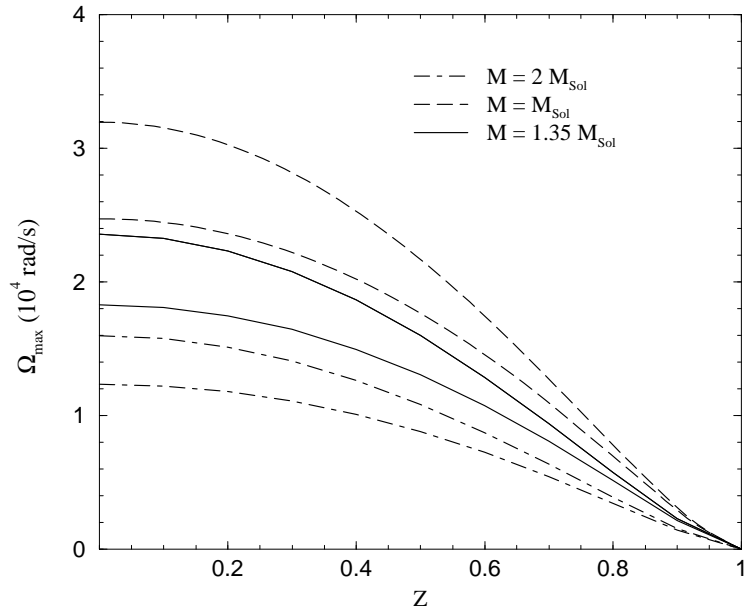}
}
\caption{Maximum angular velocity as a function of the redshift for different masses. For each mass,
there are two curves. The upper curve  is the maximum angular 
velocity (Equation (\ref{maximum angular velocity})) obtained from the perturbative metric. The lower 
curve for each mass corresponds to the analogous expression for the extended metric  
(\ref{extended metric}).}
\end{center}
\end{figure}

The emission angle has been measured indirectly for many pulsars by Kuzmin and Wu \cite{Kuzmin}, and
in their results, a strong correlation between orthogonal magnetic axes and fast millisecond pulsars
is evident. This measured correlation agrees with our results and the results of \cite{Orthogonal, Backer}, which favor orthogonal rotators.\\

On the other hand, we can assume the theory outlined above to be valid, which allows us 
to determine $\Omega_{max}$.
Indeed, we can even assume that the angular velocity of fast rotating objects is close to the
maximally allowed value. In other words we have $\Omega \approx \Omega_{max}$ which essentially,
knowing $\Omega$, predicts the redshift. In turn, we can now extract the value of the radius.
We will discuss this procedure taking realistic examples in the next section. Some examples 
of the behavior of $\Omega^{\prime}_{max}$ as a function of the gravitational redshift
are shown in figure 3. \\

There has been a considerable amount of work done related to the maximum angular velocity.
Haensel et al. \cite{Haensel1}, Glendenning \cite{Glendenning} and Koranda et al. \cite{Koranda}
have used empirical formulae together with extensive numerical calculations to provide a lower bound 
for the period of rotation. Haensel et al. \cite{Haensel1} found the minimum period to be 
$T_{min}=0.288\,ms$, assuming a mass of $1.44M\odot$. This period also depends on the minimum bound for
the redshift which was found to be $\mathcal{Z}=0.528$ \cite{Haensel1}. 
If we use eq. 
(\ref{maximum angular velocity}) (in accordance with the evidence of milli-second 
pulsars being orthogonal rotators 
we use $\theta=\pi/2$) and the parameters just mentioned we obtain $T_{min}=0.442\,ms$.
However, if we use our absolute bound (eq. (\ref{absolute max})) and the mass assumed by Haensel 
we obtain the very close value of $T_{min}=0.284\, ms$. This is just a difference of just 1.4\% 
which is rather gratifying recalling that our result has been obtained 
analytically by methods different from \cite{Haensel1}. \\

Other relevant work has been done related to the ``mass-shedding limit''  
(\cite{Latt-Prak-Masak},\cite{Friedman},\cite{Haensel}). This 
limit is conceptually  different from what we have found and corresponds 
to the limit at which the neutron 
star would break apart. Lattimer and
Prakash find for this limit
\begin{equation} 
\Omega^{(2)}_{max}(R)=1045\, (M/M\odot)^{1/2}(10\, km/R)^{2/3}\, Hz,\label{mass-shedd}
\end{equation} 
which is independent of the equation of state \cite{Lattimer} and applicable for masses not very close the maximum mass. Equation (\ref{mass-shedd}) can be used
for example, to find a limiting radius having an assumed or measured mass. While applying our results
to find a radius we will compare our findings to  
the  results when  our maximal angular velocity is replaced by the one above. Here we note that that it is not
straightforward to compare the radius independent limit (\ref{absolute max}) with (\ref{mass-shedd}) as the latter
depends explicitly on $R$. However, it is obvious that both are of the same order of magnitude.\\

The orders of magnitude reached by the maximum angular velocity are quite large, but still less than
the condition for slow rotation given by Zeldovich and Novikov \cite{Novikov}; even for the absolute
maximum (Equation (\ref{absolute max})). The condition for slow rotation ($J\ll MR_g c$) gives
\begin{equation} \label{zeld}
\Omega \ll \Omega^{(3)}_{max} \equiv \frac{5GM}{R^2 c}
\end{equation} 
which is one order of magnitude larger than our maximum upper bound when we take $R=R_g$, 
and even larger when we compare with the non-zero redshift case.
Still, because our maximum angular velocity is comparable to the Zeldovich-Novikov condition, 
especially with $R>R_g$, we will
make a better approximation for the metric in the following section.

\section{Improvements: Extended Metric}
Even though the perturbative metric (\ref{perturbative metric}) gives relevant results,   
a better approximation for the metric is given for instance in \cite{Ipser}. 
Assuming a negligible quadrupole
moment, the second approximation the metric is
\begin{eqnarray}\label{extended metric}
g_{tt}&=&-\left\{1-\frac{2 G M}{c^2 R}-\frac{1}{6}\left(\frac{G M}{c^2 R}\right)^3
-\mathcal{O}\left[\frac{G M}{c^2 R}\right]^4\right\}\\ 
g_{t\phi}&=&\frac{J \sin ^2\theta }{c^2 M}
\left\{
\frac{2 G M}{c^2 R}
+4\left(\frac{G M}{c^2 R}\right)^2
+\left(\frac{G M}{c^2 R}\right)^3
+\mathcal{O}\left[\frac{G M}{c^2 R}\right]^4
\right\}\\ \nonumber
g_{\phi \phi}&=&\frac{R^2 \sin ^2(\theta )}{c^2}\left\{
1
+\frac{2 G M}{c^2 R}
+\frac{1}{2}\left(\frac{G M}{c^2 R}\right)^2
+\mathcal{O}\left[\frac{G M}{c^2 R}\right]^3\right\} \nonumber
\end{eqnarray}
One can apply the very same procedure as before. The resulting
polynomial  to be solved to obtain the two solutions 
for the radius is of fifth order in $R$. In consequence 
we can present the solution only numerically. 
An inspection of figures 2 and 3 shown that the new results,
differ from the previous one only at very high angular velocities. The maximum angular
velocity obtained using the extended metric is slightly higher than the one obtained analytically, so
that eq. (\ref{maximum angular velocity}) for the maximum angular velocity is still a valid bound. 
However, in comparison to the
first order metric we used before, it becomes clear that the limit on the maximal angular velocity
improves, especially for the absolute limit at zero gravitational redshift.

\section{Applications}
Here we shall apply our results to some known compact objects such as Sirius B, an isolated neutron
star, and several known pulsars.\\

Sirius B is the binary companion of the very bright Sirius A and is the closest White Dwarf to earth.
Sirius B has been studied extensively, and the Gravitational Redshift
has been measured accurately with the help of the Hubble space telescope \cite{Barstow}. 
Also, since it is a binary system, its mass has been measured accurately \cite{Holberg}. 
\begin{equation}
\mathcal{Z}=0.999735 \pm 0.000015\,\,;\,\,\,M=0.984M\odot
\end{equation}
According to eq. (\ref{maximum angular velocity}), we get a minimum period of
\begin{equation}
\frac{\mathcal{T}_{min}(SiriusB)}{\sin{\theta}}=10.8\,s
\end{equation} 
This number does not change very much, had we applied the results from the extended metric.\\

%The neutron star RX J185635-3754 is the closest known neutron star
%and has been studied by Lattimer and Prakash \cite{Lattimer2}.
%The measured redshift is

The low mass X-Ray binary system EXO 0748-676 has been studied by Cottam et al. \cite{Cottam}.
They managed to measure spectral absorption lines corresponding to a redshift of

\begin{equation}
\mathcal{Z}=0.74 
\end{equation}
This together with its mass (Assumed to be $1.35M\odot$) gives
\begin{equation}
\mathcal{T}_{min}(EXO 0748-676)=9.05\times 10^{-4}s
\end{equation}
The value above has been obtained employing the extended metric (the
corresponding value resulting from the first order approximation
is $5.8 \times 10^{-4}s$). The period for this neutron star has been measured to be $22\times 10^{-3}s$\cite{villareal},
which is not too far away from the above limit given by $\mathcal{T}_{min}$.
The predicted radius for the assumed mass and measured period and redshift,
is $8.7km$.

%Perhaps it is worth mentioning that the authors of \cite{Lattimer2} 
%tried to measure periodic variations 
%in the thermal spectrum to obtain a period, however, they did not succeed. 
%It could be that this was because
%they were looking for periods of the order of seconds \cite{Lattimer2}. 
%The previous result suggests that shorter 
%periods of the order of milliseconds are possible.\\
In addition to the previous applications, our results concerning the maximum 
angular velocity, can be applied to millisecond pulsars. It is 
plausible to assume that for millisecond pulsars, the measured angular velocity is very close to 
the maximum angular velocity. With this in mind, it is possible to predict a value for 
$\mathcal{Z}$ using equation (\ref{Omax2}), and a unique value for the radius using 
(\ref{r}) or (\ref{r2}). Since
$\Omega_{max}(\mathcal{Z})$ is a decreasing function, the predicted $\mathcal{Z}$ is actually a 
maximum bound, and if the ``preferred'' radius for the neutron star is $r_2$, the predicted radius can
be thought of as a maximum bound also. Here we present a table  
with predicted redshifts and radii for 
several fast millisecond pulsars.\\

\begin{figure}
\begin{tabular}{llllll}
\hline
Pulsar            & Period (ms)     & M ($M\odot$)      &$\theta$      &$\mathcal{Z}$ & R (km) \\ \hline\hline
PSR J1748-2446ad& 1.396 \cite{Hessels}& 1.35?              &90?           & 0.834        & 20.1 \\ \hline 
PSR B1937+21   & 1.557 \cite{Ashworth}& 1.35?              & 90 \cite{Kuzmin}& 0.836        & 20.2 \\ \hline
PSR J1909-3744 & 2.95 \cite{Jacoby}& 1.438 \cite{Jacoby}& 90?            & 0.896        & 31.1 \\ \hline
PSR 1855+09     & 5.3              & 1.35?              & 90 \cite{Kuzmin}& 0.935        & 46.9 \\ \hline
PSR J0737-3039 A&22.7 \cite{Lyne} & 1.34 \cite{Lyne}   & 90?            & 0.976        & 133.6 \\ \hline
PSR 0531+21     & 33.3             & 1.35?              & 90 \cite{Kuzmin}& 0.982        & 164.4 \\ \hline
PSR B1534+12   & 37.9 \cite{Stairs}& 1.34 \cite{Stairs} & 90?            & 0.983        & 165.9 \\ \hline\hline
\end{tabular}
\caption{Predicted Maximum $\mathcal{Z}$ and radius for several pulsars.}
\label{table}
\end{figure}

From the table in can be inferred that for fast millisecond pulsars the radii are consistent with
standard neutron star models \cite{Shapiro}. However, for the slower 
millisecond pulsars, the maximum radius is slightly greater than what is predicted by most neutron 
star models, which implies
that these neutron stars are probably not rotating exactly at their maximum angular velocity. 
In such a case the given radii should be interpreted as an upper bound which comes indeed
close to the upper limit discussed in the introduction.
Using equation 
(\ref{mass-shedd}) and assuming a mass of $1.4\,M\odot$ the corresponding radius for PSR B1937+21
turns out to be $15.5\,km$ \cite{Lattimer}, $5\,km$ less than our result. For the slower 
millisecond pulsars, our results also predict slightly larger radii than those obtained by
equation (\ref{mass-shedd}).

\section{Conclusions}
We have shown that the gravitational redshift in conjunction with global results from
theoretical models can yield valuable information on the properties of 
rotating compact objects. Even though the determination of the radius in the
presence of rotation is not a unique prescription, for relatively small
angular velocities we can always opt for the lower result of the radius
determination. With increasing angular velocity,  a narrow range of possible radii
can be defined. Alternatively, assuming that the 
angular velocity of the fast spinning neutron stars is close to its maximal value,
we can either obtain a unique radius or an upper bound. The maximal
angular velocity derived in text does not depend on the radius directly, but
on the redshift which makes direct contact with existing or future observations.
The absolute limit on $\Omega$ (\ref{Omax1}) does not even depend on the redshift.
It is satisfying that both these bounds come close to the observed values for millisecond pulsar.
%Moreover, our analytical bound agrees also with results obtained in numerical calculations.
This implies that nature reaches here it maximally possible value. Another advantage of our
approach is the confirmation of the emission angle of radiation in fast rotating neutron stars.
The application of our results to existing objects clearly show that the method of using
the gravitational redshift for rotating objects is effective.\\

The important feature we would like to emphasize here is that we relate properties of Rotating 
Compact Objects to measurable quantities such as the Gravitational Redshift and the angular velocity.
This way, our approach is semi-empirical and independent of 
model details. \\

As discussed in section four, our analytical findings regarding the maximal angular velocity agree with results
obtained after numerical calculations. In this way, both, the analytical and numerical approach, corroborate each other. 

{\bf Acknowledgment}:
We would like to thank Paolo Gondolo, from the University of Utah, for 
useful discussions and carefully reading the manuscript.


\begin{thebibliography}{10}
\bibitem[22]{footnote1} This redshift factor is often defined as $(1+\mathcal{Z})^{-1}$ 
instead of just $\mathcal{Z}$. This expression was first obtained by Luminet \cite{Luminet} in a different
context.
\bibitem[24]{footnote2} Even the gravitational 
redshift of the sun has been measured (\cite{Krisher}, \cite{Snider}).
\bibitem[33]{footnote4} The highest observed spin rate is $716\;Hz$ \cite{Hessels}, 
$\mathcal{T}=1.5\times 10^{-3}s$.
\end{thebibliography}
\end{document}